\documentclass[aps,prb,twocolumn,nofootinbib,
superscriptaddress,10pt,amsmath,amssymb,floatfix]{revtex4}
\usepackage{graphicx}
\usepackage{natbib}
\usepackage{amsmath}
\usepackage{amssymb}
\usepackage[utf8]{inputenc}
\usepackage{comment}
\usepackage{fancyhdr}
\usepackage{dcolumn}
\usepackage{bm}
\usepackage{hyperref}
\usepackage{verbatim}
\usepackage{bbold}
\usepackage[T1]{fontenc}
\usepackage{amsfonts}
\usepackage{amssymb}
\usepackage{amsmath}
\usepackage{xcolor}
\usepackage{verbatim}
\usepackage[percent]{overpic}
\usepackage{soul}

\graphicspath{{pics/}}

\begin{document}

 \title{Charge separation in donor-C$_{60}$
complexes
with real-time Green's functions:\\The importance of nonlocal
correlations}
    
\author{E. Vi\~nas Bostr\"om}
\affiliation{Lund University, Department of Physics and European
Theoretical Spectroscopy Facility (ETSF), PO Box 118, 221 00 Lund, Sweden}
\email{emil.bostrom@teorfys.lu.se}
\author{A. Mikkelsen}
\affiliation{Lund University, Department of Physics and NanoLund, P.O. Box 118, 221 00
Lund, Sweden}
\author{C. Verdozzi}
\affiliation{Lund University, Department of Physics and European
Theoretical Spectroscopy Facility (ETSF), PO Box 118, 221 00 Lund, Sweden}
\author{E. Perfetto}
\affiliation{CNR-ISM, Division of Ultrafast Processes in Materials (FLASHit), Area della Ricerca di Roma 1, Via Salaria Km 29.3, I-00016 Monterotondo Scalo, Italy}
\author{G. Stefanucci}
\affiliation{Dipartimento di Fisica and European Theoretical 
Spectroscopy Facility (ETSF), Universit\`{a} di Roma Tor Vergata,
Via della Ricerca Scientifica 1, 00133 Rome, Italy}
\affiliation{INFN, Sezione di Roma Tor Vergata, Via della Ricerca
Scientifica 1, 00133 Rome, Italy}

\begin{abstract}
We use the Nonequilibrium Green's Function (NEGF) method to perform
real-time simulations of the ultrafast electron dynamics of
photoexcited
donor-C$_{60}$ complexes {\color{black}{modeled by a Pariser-Parr-Pople 
Hamiltonian}}. The NEGF results are compared to mean-field
Hartree-Fock (HF) calculations to disentangle the role of
correlations.
Initial benchmarking against numerically highly accurate
time-dependent Density 
Matrix Renormalization Group calculations verifies the accuracy of
NEGF. We then find
that charge-transfer (CT) excitons partially decay into charge 
separated (CS) states if {\em dynamical} non-local correlation
corrections
are included. This CS process occurs in $\sim 10$ fs after
photoexcitation. In contrast, the probability of exciton recombination
is almost 100\% in HF simulations. These results are largely
unaffected by nuclear vibrations; the latter become however essential
whenever level
misalignment hinders the CT process. The robust nature of our 
findings indicate that ultrafast CS driven by
correlation-induced decoherence may occur in many organic nanoscale
systems,
 but it will only be 
correctly predicted by theoretical treatments that
include time-nonlocal correlations. 
\end{abstract}

\maketitle

%
%
%

\section{Introduction}

Charge separation (CS) after photoexcitation in nanoscale
donor-acceptor (D--A) systems is the basic working principle of
organic photovoltaics.~\cite{GI.2014,SLMPS.2016,RTT.2017}
However, unraveling the mechanism responsible for the development of
a CS state poses a formidable challenge since the formation and
subsequent dissociation of the germinal charge-transfer (CT) exciton
can occur through several competing (and system-dependent) channels,
e.g., multiple CT
excitons,~\cite{Detal.2012,GMFPEBCL.2013,Setal.2014,HT.2014,WLZL.2014}
charge
delocalization,~\cite{Betal.2012,CT2012,TB.2013,H-RTB.2015,GTWI.2015,DMOB.2016}
nuclear
motion~\cite{Rozzi13,Falke14,SCPKS.2014,XZL.2015,SHSG.2015,Betal.2015,PDRFCLR.2015}
and disorder~\cite{PF2004,HK.2016}.
To characterize these processes, donor-C$_{60}$ complexes have played
an essential role as model systems and have been extensively studied
both experimentally and
theoretically~\cite{Bakulin1340,Jailaubekov13,Akimov14,Marsh10,Dowgiallo14}.

In particular, a variety of nanoscale donor systems have been coupled
to C$_{60}$ such as carbon nanotubes, pentacene and a variety of
organic complexes.
Some of the mentioned studies indicated that very fast ($<100$ fs) 
exciton dissociation can occur~\cite{Marsh10,Dowgiallo14} although, 
at the time, a direct experimental exploration of such ultrafast 
dynamics was not possible. This limitation has recently been
overcome: experimental resolution of few tens of fs  (or even less)
is now readily available for surface studies, giving direct access to
the sub 100 fs regime~\cite{Marsell15,Spektor1187,Shibuta16}. 
Even so, for a comprehensive interpretation of experiment, a 
comparison with numerical simulations
of the electron and nuclear  dynamics is needed.

Statistical approaches like Time-Dependent (TD) Density Functional 
Theory\cite{RG.1984,onida2002electronic,Ubook,maitra2017charge}
(DFT) have been central in efforts to 
model charge transfer dynamics at the fs timescale. TD-DFT can in 
principle treat all electronic excitations (Frenkel excitons,  CT 
excitons, CS states, etc.) on equal footing through the Kohn-Sham
equations for the electron density. The performance of TD-DFT
crucially relies on the quality of the {\em exchange-correlation
potential}, which is often approximated by a space- and time-local
functional of the density. As we shall see, a time-local
approximation is not always reliable.

An approach that includes in a natural way space-time nonlocal 
correlations is the NonEquilibrium Green's Function (NEGF) 
theory.\cite{KBbook,SvLbook,BBbook} Here the accuracy of the results 
relies on the quality of the so called {\em correlation self-energy} 
$\Sigma$. If $\Sigma=0$ then NEGF is equivalent to the 
(time-local) TD Hartree-Fock (HF) theory.
Since approximations to $\Sigma$ can easily be generated and 
systematically improved by means of diagrammatic expansions, 
NEGF provides a natural framework to go beyond TD-HF, 
thus including space- and time-nonlocal correlation 
corrections to the electron dynamics.

In this work we highlight the crucial role that nonlocal 
correlations can have in the CS process of a prototypical D--A
system. 
The NEGF method is initially benchmarked in a
one-dimensional D--A model system against numerically accurate
results 
from the TD Density Matrix Renormalization 
Group~\cite{White92,DeChiara2008} (tDMRG), finding excellent
agreement. 
In this assessment we also consider HF dynamics, as a paradigm
of mean-field treatments, to help disentangle the
role of correlations.
Then, we consider a [HOMO+LUMO]--C$_{60}$ dyad 
{\color{black}{modeled by a Pariser-Parr-Pople (PPP) 
Hamiltonian~\cite{PariserParr1953,Pople1953} 
with a single $p_{z}$ orbital per Carbon,}}
and show that the self-energy can qualitatively change the mean-field
dynamics at clamped nuclei, from almost 100$\%$ recombination of the 
CT exciton in HF to substantial CT and subsequent CS in correlated
NEGF simulations. We find that the CS can happen already within 
10~fs after the onset of photoexcitation.
We further show that the results are largely unaffected by nuclear 
vibrations, which do instead play a pivotal role 
whenever the misalignment of the equilibrium levels prevents the
formation of 
a CT exciton.

The paper is organized as follows: in Section~\ref{sec:model} we
describe the
system and the inherent Hamiltonian. In Section~\ref{sec:methods} we
present the
theoretical approach; the choice of system parameters is discussed in
Section
\ref{sec:ground_state}; here an analysis of the possible choices for 
the initial state is also addressed. In Section~\ref{sec:tdmrg}, we 
benchmark our Green's function approach against an exact 
solution for a paradigmatic 1D model. 
The role of electronic correlations for the donor-C$_{60}$ real-time
dynamics is presented in Section~\ref {sec:nonequilibrium}. Finally,
in Section~\ref{sec:nuclear}
the significance of the electron-phonon interactions is considered.
Our conclusions and outlook are provided in
Section~\ref{Conclusions}.

\section{Donor-C$_{60}$ Hamiltonian}\label{sec:model}
\begin{figure}[tbp]
 \includegraphics[width=1\columnwidth]{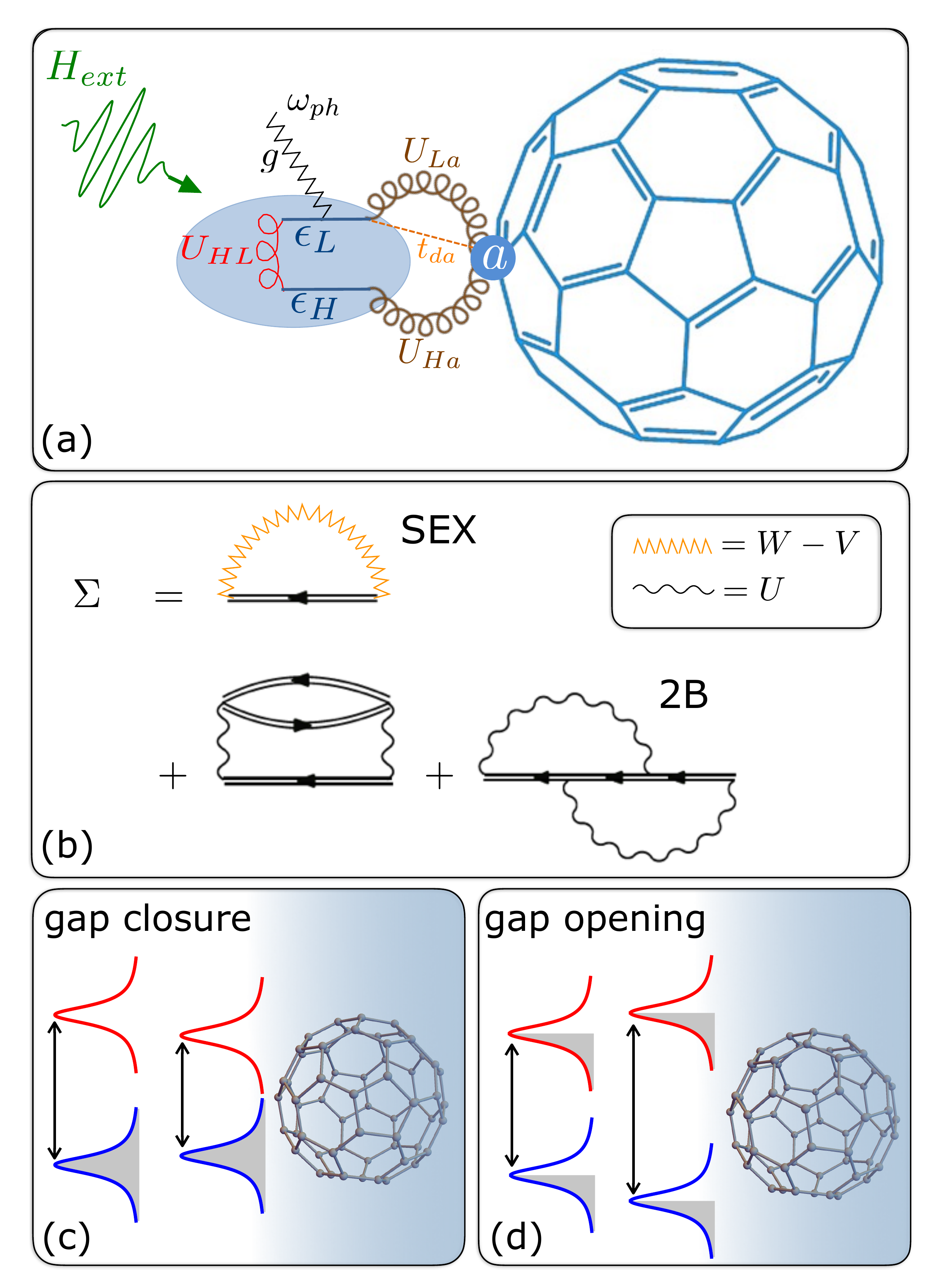}
 \caption{\small (a) Sketch of the model and experiment. The
nomenclature used for 
 the terms in the Hamiltonian is also shown. 
  (b) Self-energy approximation consisting of the SEX 
 diagram (first term)  and the 2B diagrams (second and third term). 
 Gap closure for a filled HOMO and empty LUMO (c) and gap opening for 
 a half-filled HOMO and LUMO (d) due to polarization effects induced 
 by C$_{60}$ (or any polarizable molecule or surface).}
 \label{sigmadiag}
\end{figure}

We consider a dyad consisting of a donor molecule coupled to a 
C$_{60}$ acceptor, see Fig.~\ref{sigmadiag}(a),
with Hamiltonian
\begin{align}\label{eq:ham}
\hat{H}(t) &= \hat{H}_d + \hat{H}_a + \hat{H}_{da} + \hat{H}_{e-ph} +
\hat{H}_{ext}(t)
\end{align}
where $\hat{H}_{d/a}$ describes the donor/acceptor molecule 
and $\hat{H}_{da}$ the D--A coupling. The interaction with nuclear 
vibrations is contained in $\hat{H}_{e-ph}$ whereas the interaction
with  
external electromagnetic fields (the photoexcitation indicated in
Fig.~\ref{sigmadiag}(a)) is accounted for by $\hat{H}_{ext}(t)$.

To facilitate the analysis we use a minimal 
paradigmatic description of the donor molecule, i.e., 
only two active orbitals (HOMO and LUMO)
denoted by $H$ and $L$:
\begin{align}\label{eq:ham_d}
\hat{H}_d = \epsilon_H \hat{n}_H + \epsilon_L \hat{n}_L + +
U_{HL}\hat{n}_H \hat{n}_L.
\end{align}
Here $\epsilon_X$ is the energy of the $X=H,L$ level, $\hat{n}_X =
\hat{n}_{X\uparrow} + \hat{n}_{X\downarrow} =
\hat{c}^\dagger_{X\uparrow}\hat{c}_{X\uparrow}
+ \hat{c}^\dagger_{X\downarrow}\hat{c}_{X\downarrow}$ is the
occupation
operator of the same level and $U_{HL}$ is the Coulomb interaction 
between electrons on different levels. 
For the C$_{60}$ acceptor we use a single $p_{z}$-orbital per 
carbon and write
\begin{align}\label{eq:ham_a}
\hat{H}_a &= -\sum_{ij\sigma}
t_{ij}\left(\hat{c}_{i\sigma}^\dagger \hat{c}_{j\sigma} + h.c.\right)
+  \sum_i
V\hat{n}_{i\uparrow}\hat{n}_{i\downarrow}, \nonumber \\
&\quad + \sum_{i\neq j} \lambda V\frac{
\hat{n}_{i} \hat{n}_{j}}{\sqrt{\lambda^2+d_{ij}^2}}  
\end{align}
where $t_{ij}$ and $d_{ij}$ are, respectively,
the hopping amplitude and  the distance between carbons $i$ and  
$j$. We take $t_{ij} =
t$ for two atoms belonging to the same pentagon and $t_{ij} = t'$ for
two atoms in different pentagons.
In Ohno's parametrization~\cite{Ohno64} of the interaction 
(last term of Eq.~(\ref{eq:ham_a}))
$\lambda$ accounts for the screening from the frozen shells.

We consider a donor with the HOMO orbital located far enough
from the C$_{60}$ $p_{z}$-orbitals  that we can neglect their 
overlaps. The LUMO orbital is instead located in the neighbourhood of 
the acceptor and has a 
substantial overlap with only one $p_{z}$-orbital (labelled by 
the index $i=1$); let $t_{da}$ be the corresponding hopping 
integral.
Accordingly, the Hamiltonian for the D--A
interaction is described by
\begin{align}\label{eq:ham_da}
H_{da} &= -t_{da} \sum_\sigma \left(\hat{c}_{L\sigma}^\dagger
\hat{c}_{1\sigma} +
h.c.\right)  \nonumber \\
&+  \sum_{i} \left[ U_{Hi}\left(\hat{n}_H - 2\right) +
U_{Li}\hat{n}_L\right] \left(\hat{n}_i -1\right), 
\end{align}
where $U_{Xi}$ is the strength of the Coulomb repulsion between an 
electron on $X=H,L$ and an electron on carbon 
$i${\color{black}{~\cite{TR.2009,MTKSvL.2012}.
The interaction correctly vanish in the charge neutral ground state 
(2 electrons on the HOMO and 1 electron per C atom on the C$_{60}$) 
of the subsystems  D and A at  infinite distance ($t_{da}=0$).}}
The index $i$ could in principle run across all the $60$ atoms of the
acceptor. In the actual calculations we considered only 
atom $i=1$ and its nearest neighbors.

To demonstrate the robustness of the correlated results we 
also investigate the effects of harmonic nuclear motion on the
electron 
dynamics. We model the electron-phonon interaction as 
\begin{align}\label{eq:ham_eph}
\hat{H}_{e-ph} = \frac{p^2}{2M} + \frac{1}{2}M\omega_{ph}^2 x^2 +
g \hat{n}_L x,
\end{align}
which describes  a 
Holstein-like phonon~\cite{Holstein59} coupled to the charge density
of the LUMO level.

An external light field is used to promote an
electron from the HOMO to the LUMO level.
We therefore restrict the 
light-matter interaction to the donor site
\begin{align}\label{extham}
\hat{H}_{ext}(t) = {\color{black} 2A}\cos(\omega t)\sum_\sigma
\left(\hat{c}_{H\sigma}^\dagger \hat{c}_{L\sigma} + h.c. \right).
\end{align}
The action of $\hat{H}_{ext}$ starts at $t=0$ and ends at some time 
$t_{\rm off}$
when the LUMO population (initially zero) has become close to unity.
In this work we consider a frequency $\omega = \epsilon_L - 
\epsilon_H$ close to resonant absorption.

%
%
%

\section{Method}\label{sec:methods}

As the $p_z$-orbitals are half-filled we use the statically 
screened exchange (SEX) approximation to treat the bare interaction
$V_{ij}=\lambda V/\sqrt{\lambda^2+d_{ij}^2}$ on the
C$_{60}$.\footnote{Our treatment corresponds to the static Coulomb
hole plus screened exchange (COH-SEX) approximation first introduced
by Hedin~\cite{Hedin65}. However, the Coulomb hole part only gives a
constant shift of the diagonal of the self-energy, that can be
absorbed by redefining the single-particle energies.}
Thus, the first contribution to $\Sigma$ is given by the first 
diagram in Fig.~\ref{sigmadiag}(b) where (in matrix notation)
\begin{align}\label{eq:w_int}
W= \frac{V}{1-V\chi_0},
\end{align}
and $\chi_0$ is the zero-frequency  response 
function~\cite{SvLbook}
\begin{align}\label{eq:chi0}
\chi_{0,ij} = 2\sum_{n\bar{n}}
\frac{\phi_{n}(i)\phi_{\bar{n}}(i)\phi_{n}(j)\phi_{\bar{n}}(j)}{\epsilon_n-\epsilon_{\bar{n}}}.
\end{align}
In Eq.~(\ref{eq:chi0}) $\phi_n$ and $\epsilon_n$ are the HF orbitals 
and energies of the isolated C$_{60}$, and the indices $n$ and
$\bar{n}$ run over occupied and
unoccupied states respectively. We emphasize that the screening
described by $W$ 
comes solely from the valence electrons and it is therefore of
different physical origin than that described by $\lambda$ (which is
due 
to the frozen electrons not included explicitly in our description).
Hence, no double counting is involved. 

The electron-hole attraction experienced by the CT exciton is
responsible for the 
renormalization of the HOMO-LUMO gap, an effect missed by HF 
and even by Hartree+SEX. The gap tends to close in 
equilibrium (filled HOMO and empty LUMO), see Fig~\ref{sigmadiag}(c),
whereas it tends to open in 
the photoexcited donor (half-filled HOMO and LUMO), see 
Fig~\ref{sigmadiag}(d) and Supporting 
Information. The gap closure in the equilibrium case is of relevance
in, e.g., 
estimating the conductance of a molecular 
junction.~\cite{NHL.2006,TR.2009,KF.2008,MTKSvL.2012} The gap opening 
in the photoexcited case is instead of relevance to CS since 
the formation of a CT exciton strongly depends upon level
alignment. Both types of gap renormalization are captured  by the
second-Born (2B) approximation to the self-energy. We therefore 
add to the SEX diagram
the last two diagrams in Fig.~\ref{sigmadiag}(b) where $U$ is the 
interaction in Eqs.~(\ref{eq:ham_d}) and  (\ref{eq:ham_da}).
 
In the NEGF method the total self-energy 
$\Sigma$ is used to generate the so called Kadanoff-Baym 
equations\cite{KBbook,SvLbook,BBbook,KB.2000,DvL.2007,MSSvL.2008,MSSvL.2009,vFVA.2009,MHCV2014}
(KBE) for the 
Green's function. The KBE are subsequently converted into an 
integro-differential equation for the single-particle density matrix 
using the Generalized Kadanoff-Baym Ansatz~\cite{LSV.1986,HJbook} 
(GKBA), which has proven to drastically reduce the computational cost 
without loosing accuracy~\cite{HSB.2014,SB.2014} and it has recently 
been implemented in the context of quantum transport,~\cite{Latini14}
equilibrium absorption,~\cite{PPHS.2011}  transient 
absorption,~\cite{PUvLS.2015,PSMS.2015,SDCMCM.2016,Petal.2016}
carrier dynamics in 
semiconductors~\cite{SM2015,PSMS.2016} and many-body 
localization.~\cite{BR.2014,LB.2017} The derivation of 
the GKBA equation is provided in the Supporting Information.

As for the effects of nuclear vibrations we use a 
NEGF-Ehrenfest 
approximation,\cite{BMKVA.2016,BalzerBonitz} i.e., 
we replace the momentum and position operators
with their expectation values in the GKBA equation and 
evolve the nuclear degrees of freedom via 
the Ehrenfest
dynamics.

%
%
%

\section{On the initial state}\label{sec:ground_state}

\begin{figure}[!thb]
 \includegraphics[width=\columnwidth]{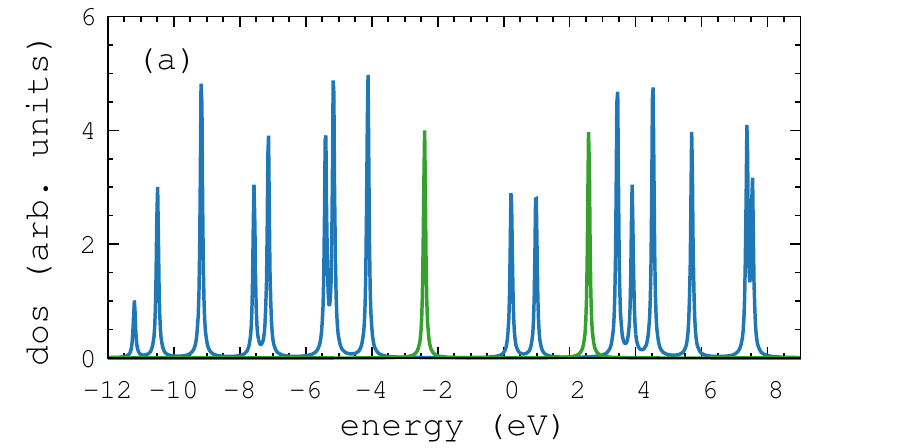}
 \includegraphics[width=\columnwidth]{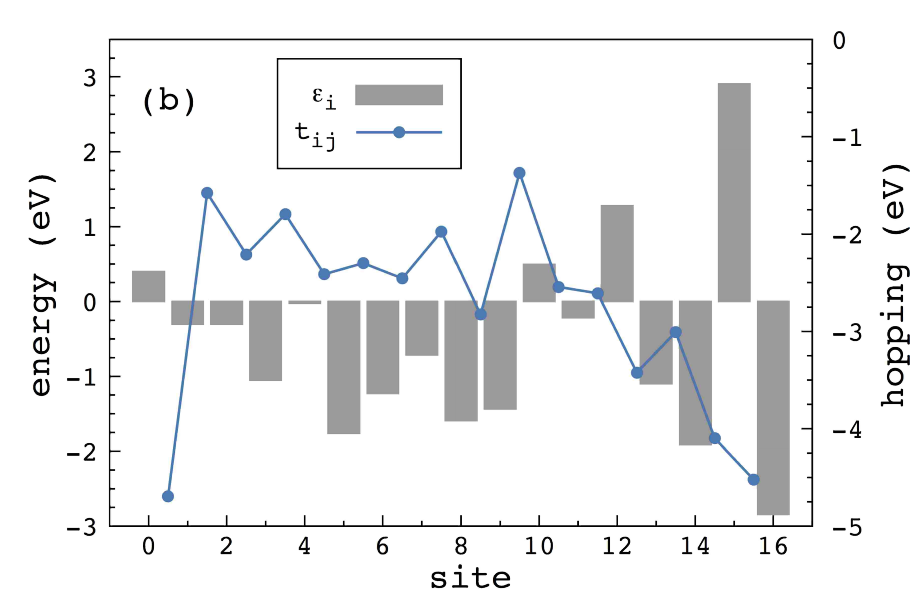}
 \includegraphics[width=\columnwidth]{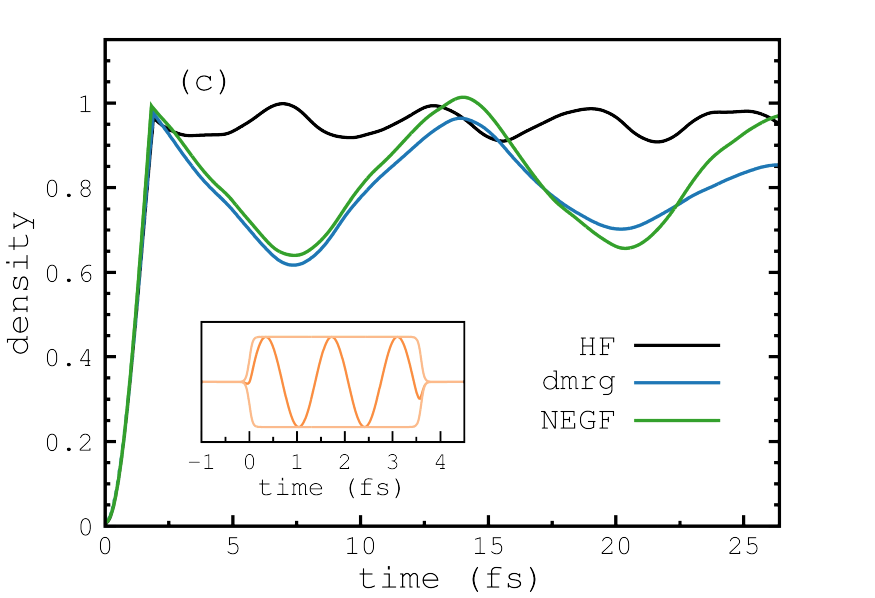}
 \caption{\small (a) HF Density of states for the
coupled donor-C$_{60}$ system, with blue peaks denoting C$_{60}$
levels and green peaks the HOMO and LUMO of the donor. (b) 
Parameters of the one-dimensional Lanczos chain of the noninteracting 
C$_{60}$. Gray bars indicate the values 
of the onsite energies $\epsilon_i$ whereas blue dots the nearest 
neighbour hopping integrals $t_{ij}$. (c) Electron 
occupation of the LUMO level for the donor coupled to the Lanczos
chain using HF, NEGF and tDMRG. {\color{black}{The inset shows the profile of the 
laser pulse along with its rectangular evelope.}}}
 \label{fig:density_tdmrg}
\end{figure}

In accordance with the
literature~\cite{Willaime93,Aryanpour10,Schmalz11}
the values of the hopping
integrals have been set to $t = 2.5$~eV and $t'  = 2.7$~eV.
The strength of the interaction on the C$_{60}$ takes the value 
$V = 8$~eV with $\lambda =0.7$~\AA\cite{Aryanpour10,Schmalz11}.
{\color{black} The D--A coupling is smaller than the C$_{60}$
parameters since
no chemical bond is formed between the molecules.
We have found that in this regime, i.e. as long as $t_{da}$ is the
smallest energy scale in the system, the results 
do not change qualitatively for different couplings. In the following
we set the hopping integral $t_{da} =
0.4$~eV.} The D--A interaction strength is taken as $U_{Li} =
U_{La}$ and $U_{Hi} = U_{Ha}$ for $i=1$, $U_{Li} =
U_{La}/2$ and $U_{Hi} = U_{Ha}/2$ for the nearest 
neighbours of $i=1$ and zero otherwise, where $U_{La} = 2U_{Ha} = 
2$~eV ($U_{La}>U_{Ha}$ since the HOMO is further away than the LUMO 
from the C$_{60}$). 
The bare energies of the HOMO and LUMO levels are
$\epsilon_{H}=-2.4$~eV, 
$\epsilon_{L}=0.6$~eV whereas the HOMO-LUMO repulsion is
$U_{HL}=1$~eV. 
{\color{black}{Unless explicitly stated, all results refer to this set
of parameter 
values.}}
The resulting {\em equilibrium} HF density of states is shown in
Fig.~\ref{fig:density_tdmrg} (top panel), and is in good qualitative
agreement
with both experiment and other theoretical
results~\cite{Dresselhaus96,You94,Braga91}.

The first step in a CT process is the photoexcitation of the electron
from the initially occupied HOMO. 
{\color{black}{Due to absence of spin symmetry breaking terms, 
the occupations $n_{X\sigma}$ ($X=H,L$) remain independent of 
$\sigma$ during the entire time evolution.}}
In photoemission 
a common way of treating the photoexcitation is to employ the sudden
approximation~\cite{Kuleff14}, wherein the HOMO hole is assumed to
be created instantaneously. The rationale behind this approximation
is that 
the time it takes for the electron to escape can be made arbitrarily 
short by using photons of high frequency. The sudden approximation
applied to the 
HOMO-LUMO transition would yield the initial state 
$|\Psi^{(1)}_{\sigma}\rangle =
\hat{c}^\dagger_{L\sigma}\hat{c}_{H\sigma}|\Psi_0\rangle$, where
$|\Psi_0\rangle$
is the ground state of the system.
However, for the HOMO-LUMO transition,
the time to transfer an electron is ultimately bounded by the inverse 
of the Rabi frequency. 
From the exact solution of the two-level system driven by the 
perturbation in Eq.~(\ref{extham})
 one finds that the 
state with a half-filled HOMO and LUMO reads 
$|\Psi^{(2)}\rangle =
\frac{1}{4}\left(c_{L\downarrow}^\dagger
-i\hat{c}_{H\downarrow}^\dagger\right)\hat{c}_{H\downarrow}\left(\hat{c}_{L\uparrow}^\dagger
-i\hat{c}_{H\uparrow}^\dagger\right)\hat{c}_{H\uparrow}|\Psi_0\rangle$.
We thus see that $|\Psi^{(2)}\rangle$ is a linear combination of the 
ground state $|\Psi_0\rangle$, the singly excited states 
$|\Psi^{(1)}_{\sigma}\rangle$ and the doubly excited 
state 
$\hat{c}_{L\downarrow}^\dagger
\hat{c}_{H\downarrow}\hat{c}_{L\uparrow}^\dagger
\hat{c}_{H\uparrow}|\Psi_0\rangle$. We have compared 
the dynamics obtained using
$|\Psi^{(1)}_{\sigma}\rangle$ and $|\Psi^{(2)}\rangle$ as initial 
states, as well as direct excitation using an
external field, and found scenarios where the sudden approximation
fails dramatically. In this paper we exclusively use an external
field to initiate the dynamics. {\color{black} We take the amplitude in 
Eq.~(\ref{extham}) to be  $A=0.3$~eV, which given a typical HOMO-LUMO
dipole $d = 1\div 10$ a.u. corresponds to a laser intensity of $I =
4.2\times(10^{10}\div 10^{12})$ W/cm$^2$. A discussion on the 
dependence of the excitation
pulse and on the differences with the sudden approximation 
is given in the Supporting Information.}

%
%
%

\begin{figure*}
\includegraphics[height=5cm]{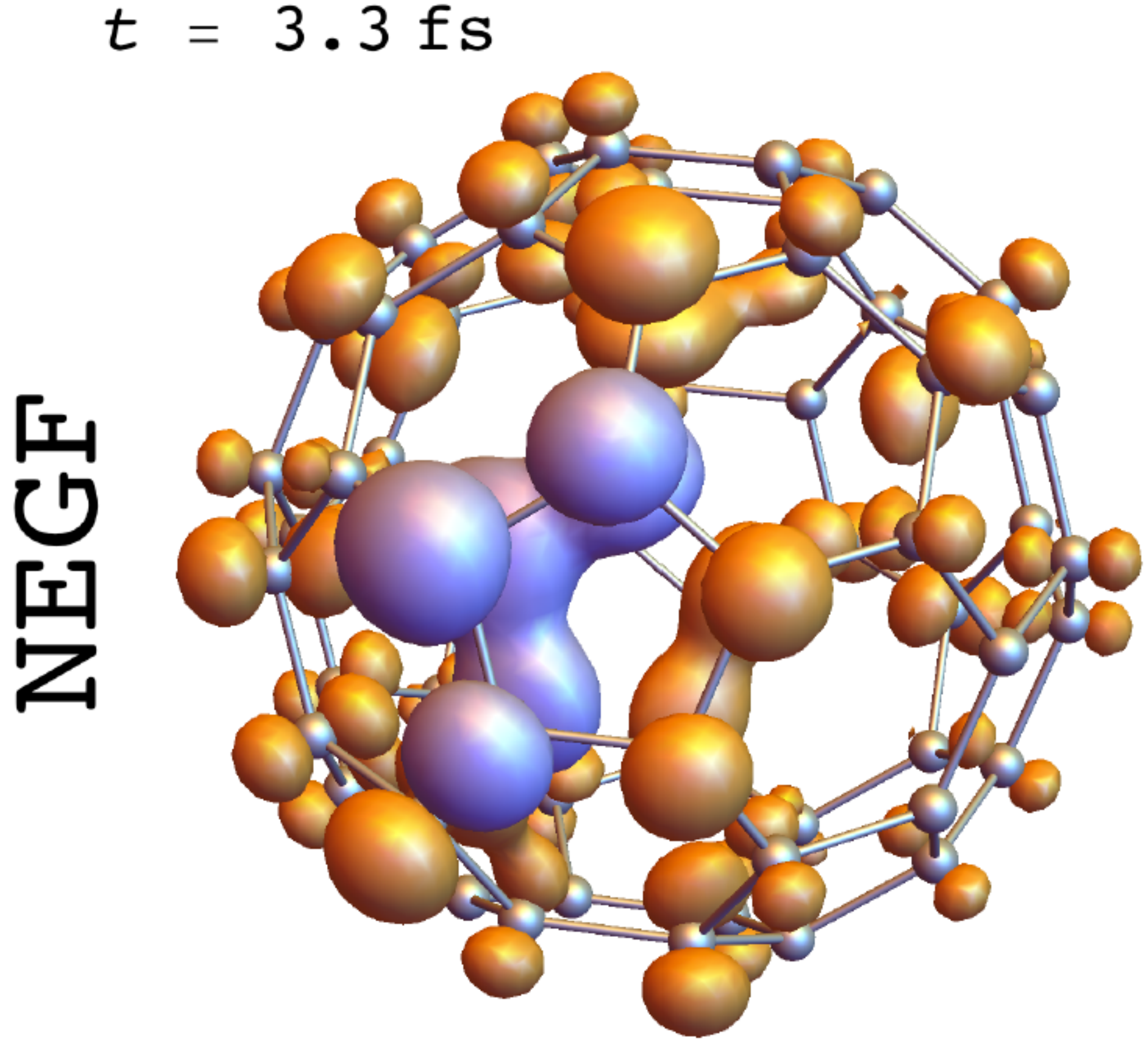}
\includegraphics[height=5cm]{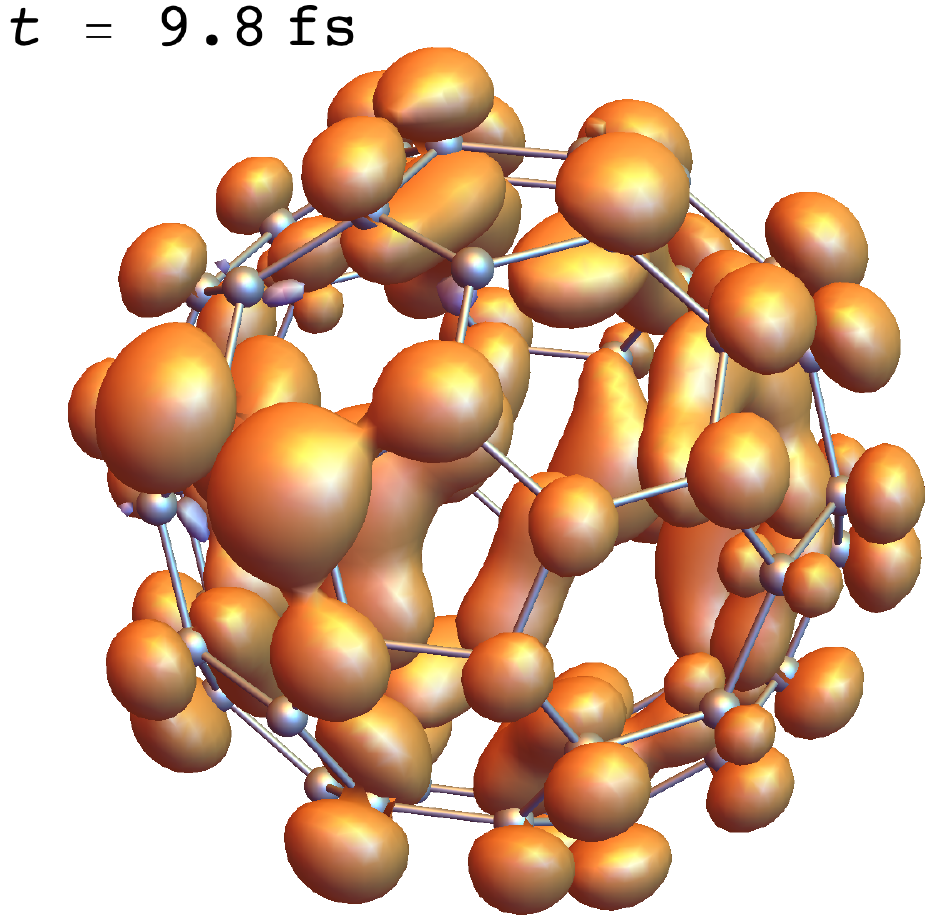}
\includegraphics[height=5cm]{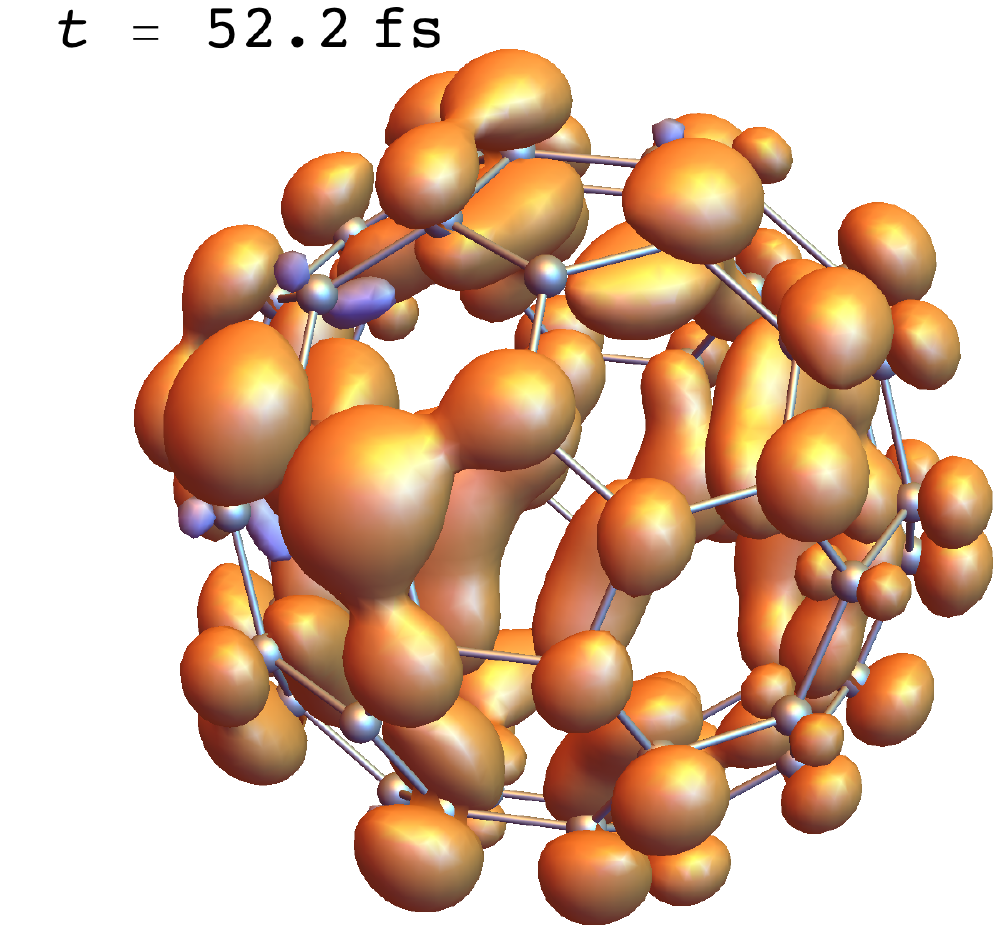}
\\
\vspace{0.5cm}
\includegraphics[height=5cm]{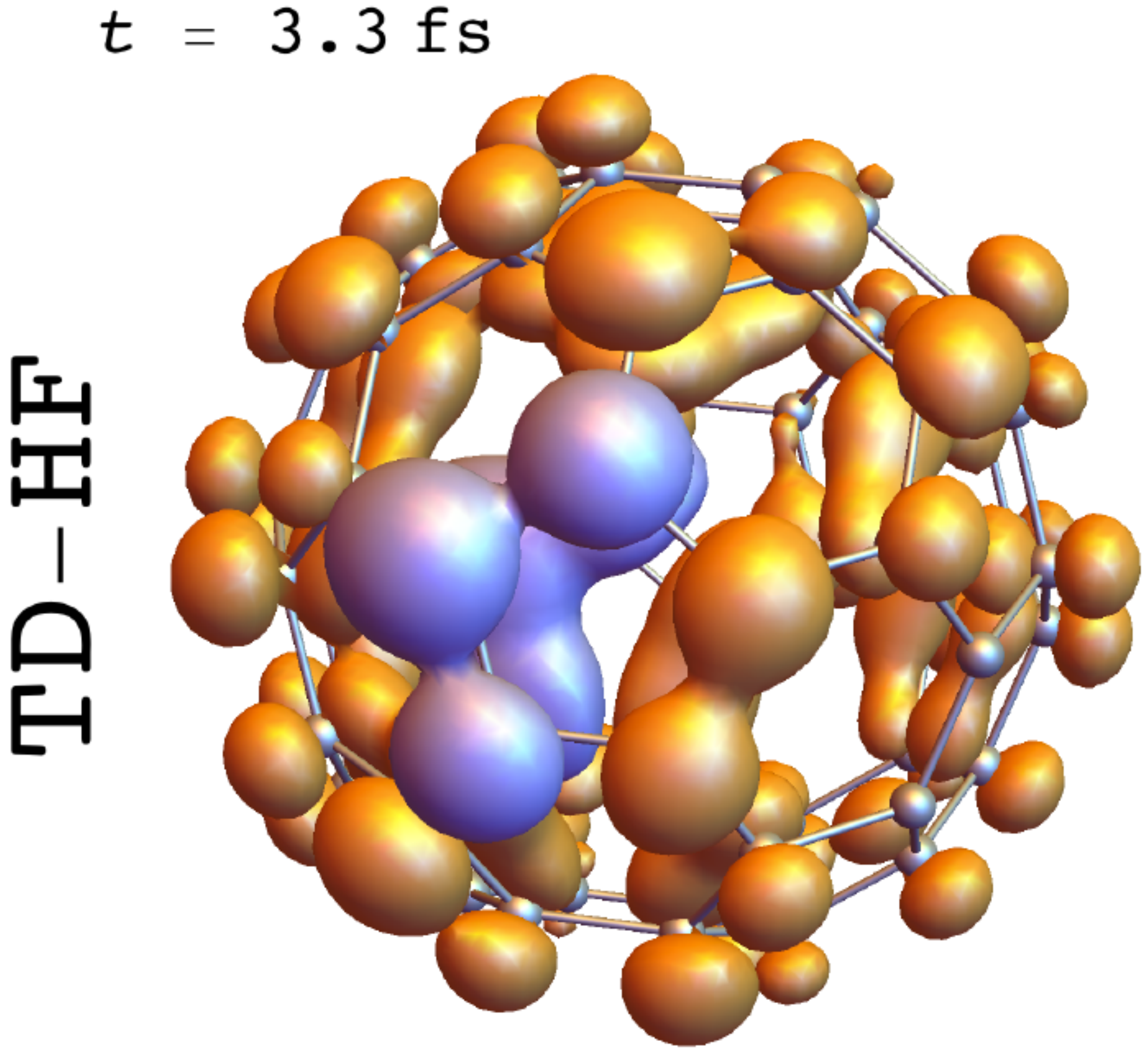}
\includegraphics[height=5cm]{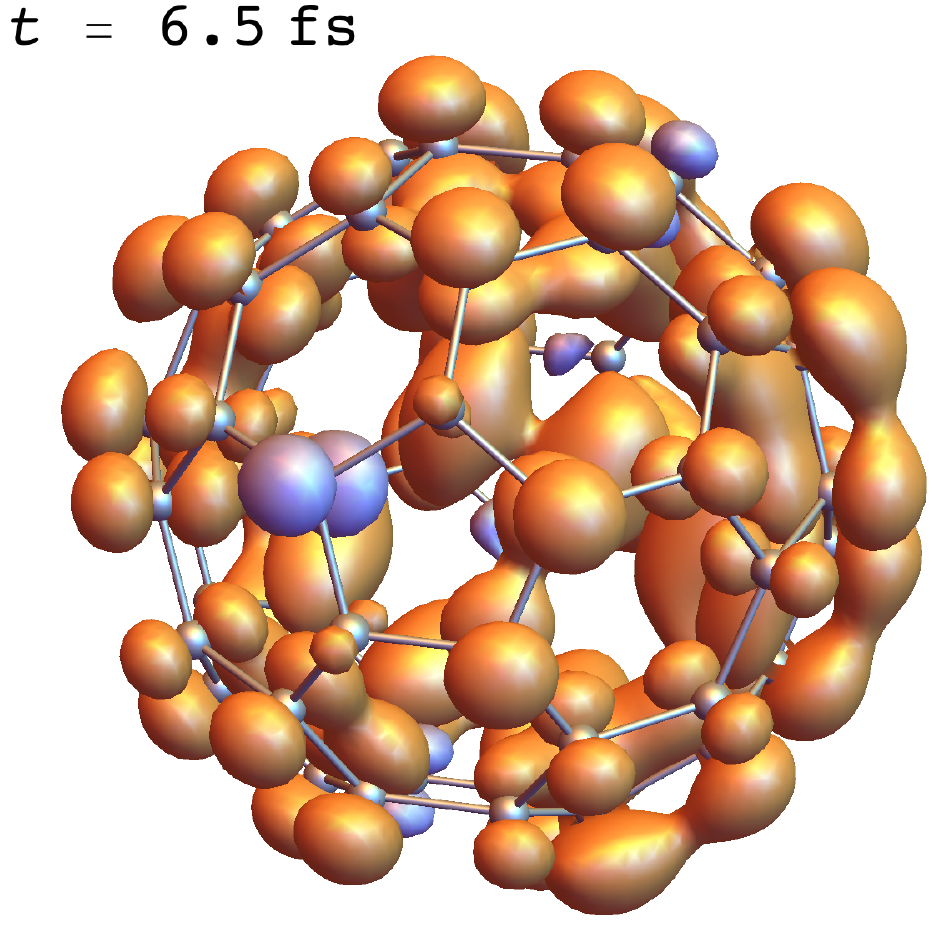}
\includegraphics[height=5cm]{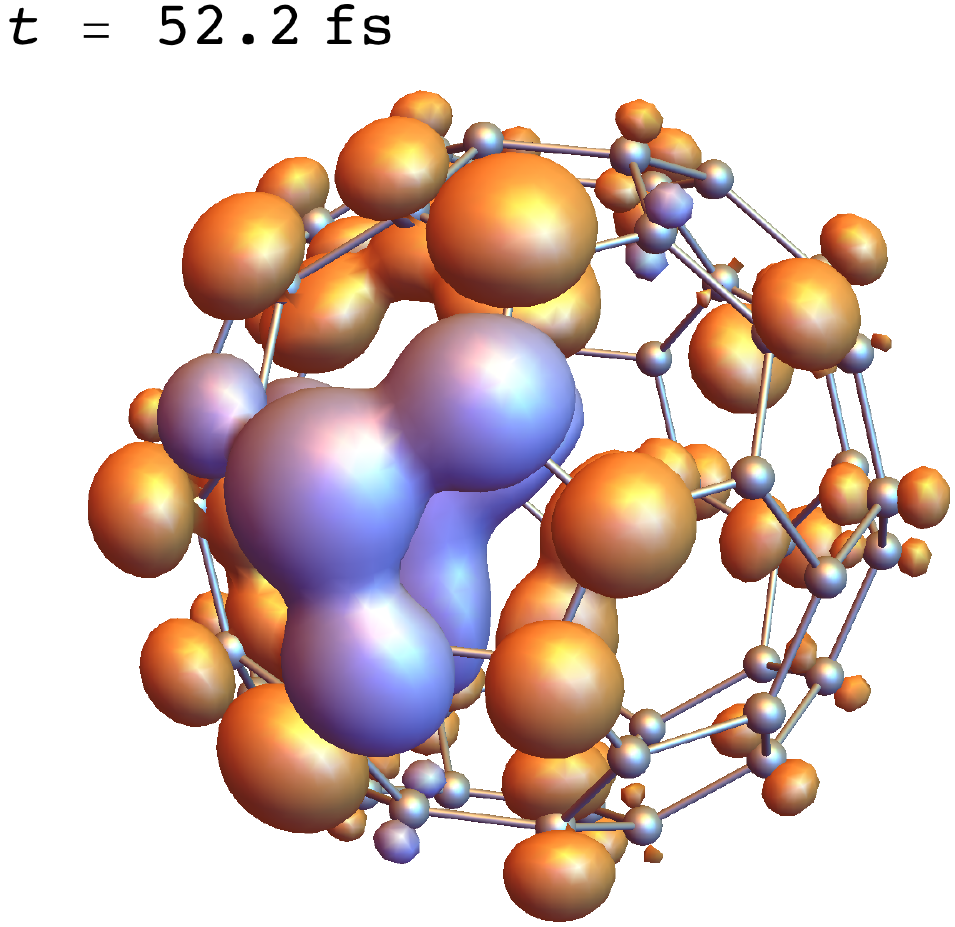}
\caption{\small Charge dynamics in NEGF (upper row) and TD-HF (lower
row) 
for the C$_{60}$ acceptor. In each row, the three time snapshots 
depict the excess density as defined in Eq.~(\ref{excessn}) (blue 
for $\Delta n<0$ and orange for $\Delta n>0$). {\color{black}{With
reference to 
Fig.~\ref{fig:densities_c60_hf}a--b,}} they respectively 
correspond to the time of maximum photo-charging of the LUMO 
($t=3.26$ fs), the first time the LUMO reaches minimum density (at 
$t=9.78$ fs for NEGF and $t=6.52$ fs in TD-HF)   
and its later occupation at $t=52.16$ fs.}
 \label{fig:densities_c60_2b}
\end{figure*}


\begin{figure*}
\includegraphics[scale=1]{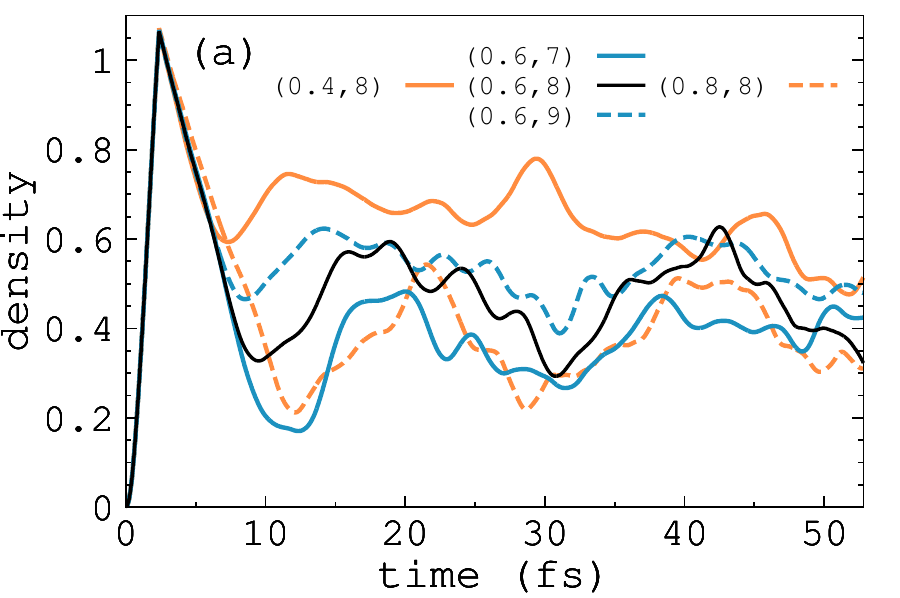}\hspace*{0.4cm}%
\includegraphics[scale=0.12]{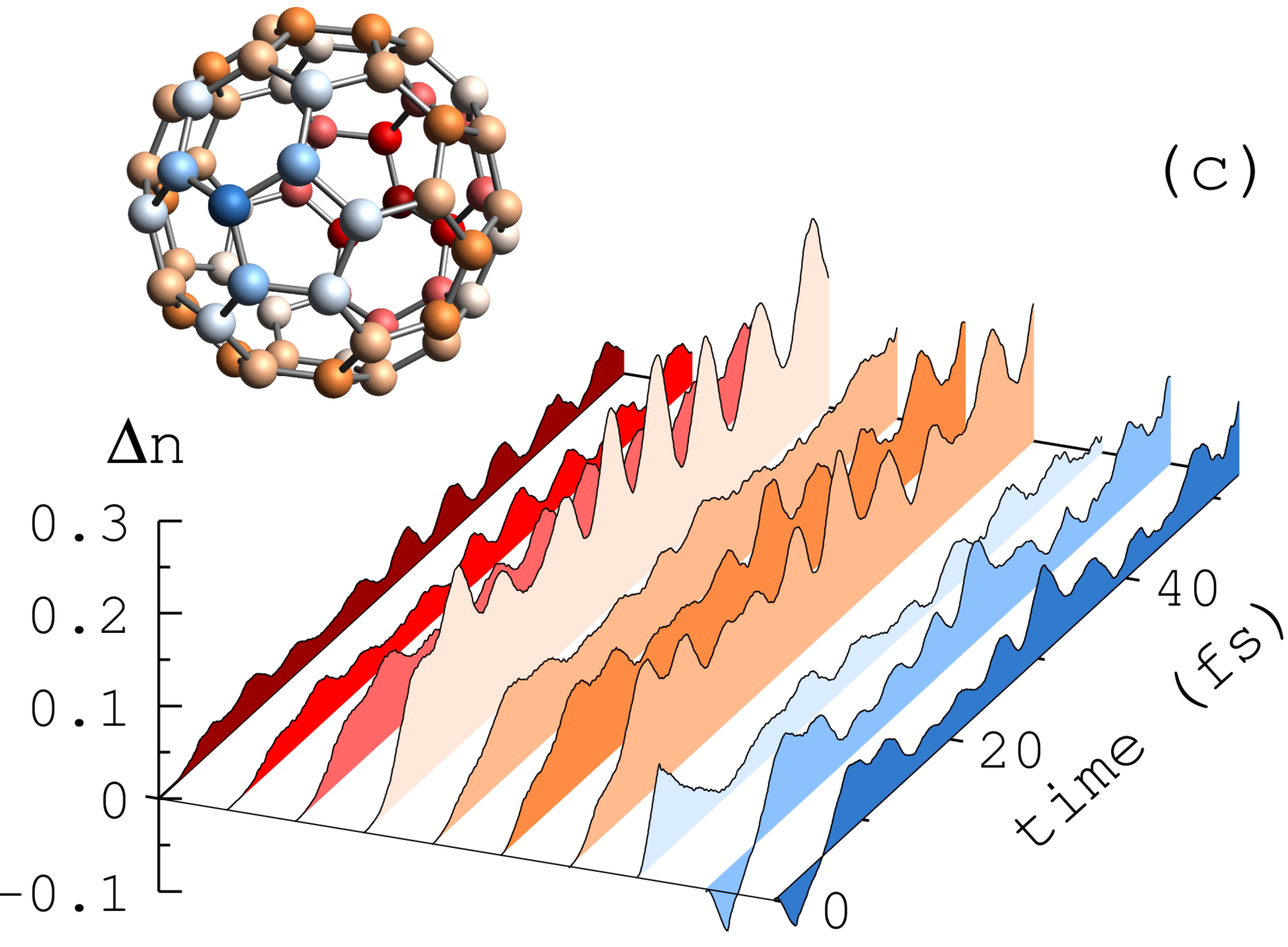}

\vspace{0.5cm}
\includegraphics[scale=1]{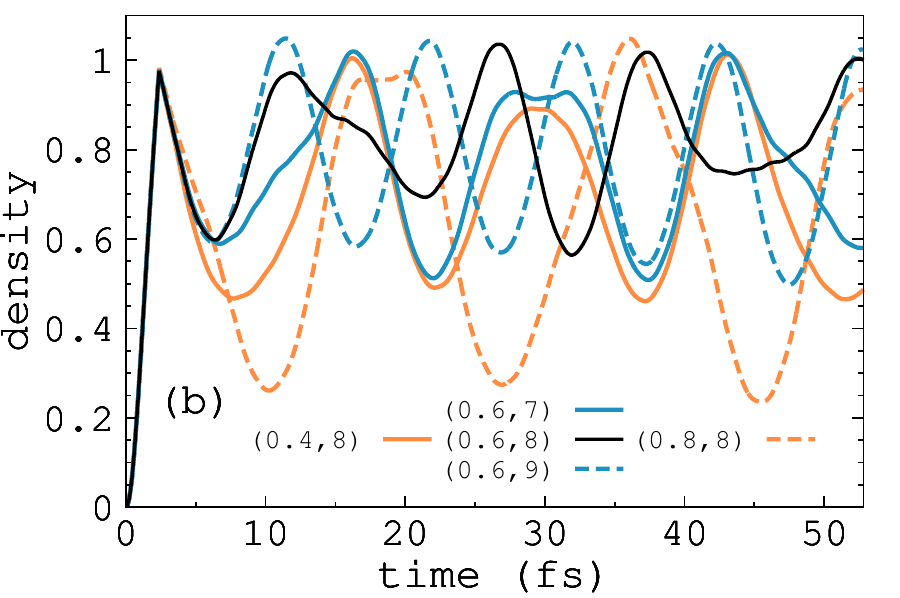}\hspace*{0.4cm}%
\includegraphics[scale=0.12]{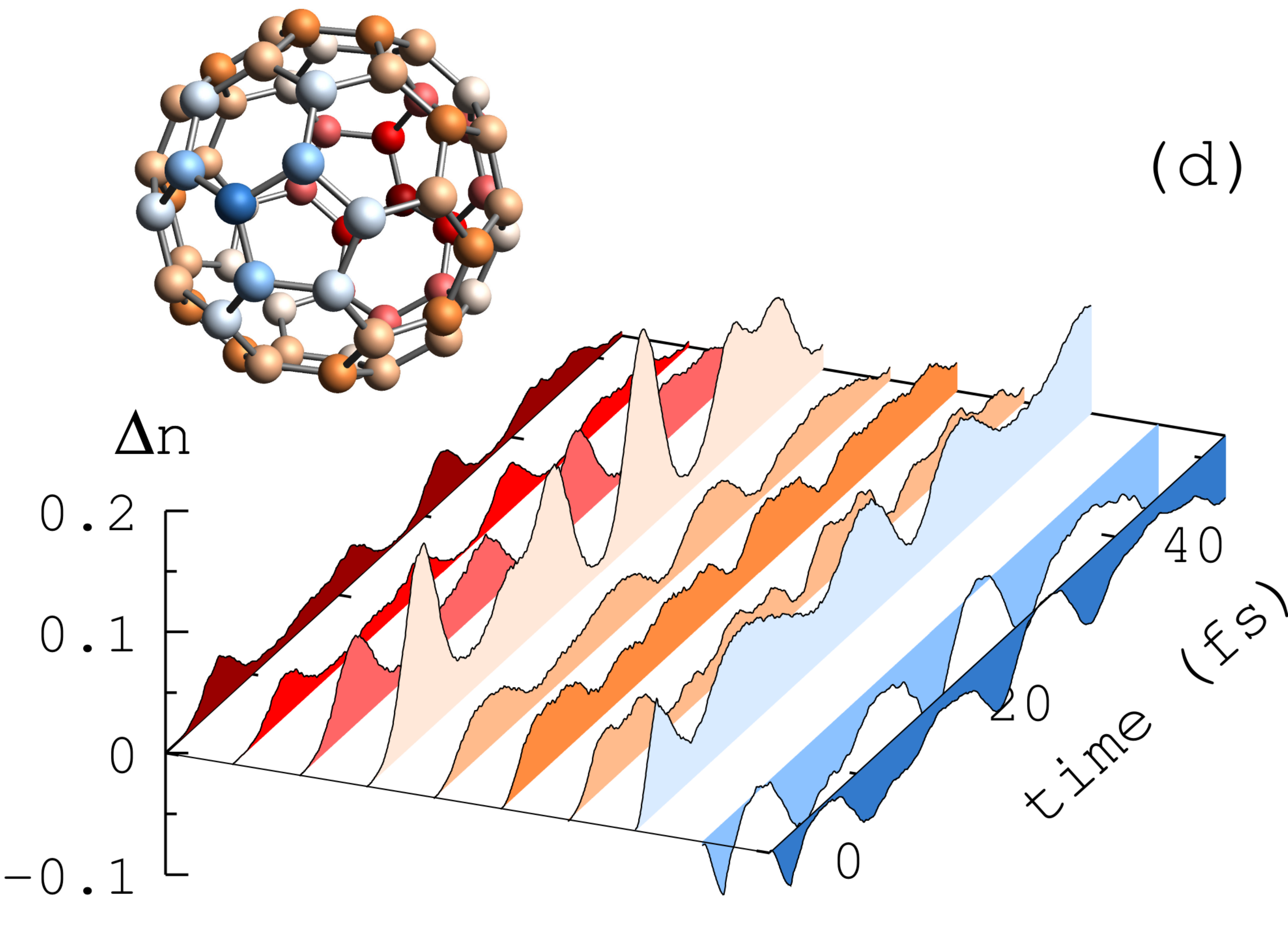}
\caption{\small NEGF (a) and TD-HF (b) results for the
LUMO density for different values (in units of eV) of 
{\color{black}{
$(\epsilon_L,V)=(0.4,8),\,(0.6,7),\;(0.6,8),\;(0.6,9),\;(0.8,8)$ 
[the default parameters $(\epsilon_L,V)=(0.6,8)$ give the black curve]}}. 
The excess density in the C$_{60}$ shells as a function of time
is shown in panels (c) and (d) for NEGF and TD-HF, respectively.
Atoms of the same shells have the same color, as indicated in the
C$_{60}$ cluster.}
\label{fig:densities_c60_hf}
\end{figure*}

%
%
%

\section{Benchmarks with tDMRG}\label{sec:tdmrg}
The accuracy of NEGF results is discussed in this Section by direct 
comparison with tDMRG  on a simplified system.
Since tDMRG is a numerically highly accurate scheme for
one-dimensional models, we map
the donor-C$_{60}$ Hamiltonian onto a linear chain. For the 
noninteracting C$_{60}$ this is achieved using the
Lanczos method~\cite{Lanczos50}, and results in a $17$ site chain
with different nearest neighbor hopping and on-site
energies \footnote{
Use of the Lanczos technique to reformulate a problem via a 1D effective Hamiltonian, is a quite common procedure, notably in theoretical surface science. For a recent application in quantum transport, see e.g. [\onlinecite{Olevano10}]}. The 
parameters of the Lanczos chain are shown in
Fig.~\ref{fig:density_tdmrg} (middle panel). 
We then include interactions
between the LUMO level and the first site of the 
chain (tDMRG is most effective for short-range interactions) of 
strength $U_{La} = 0.8$ (here $U_{Ha} = 0$). 

We perform TD simulations for the monochromatic driving of 
Eq.~(\ref{extham}) switched on {\color{black} approximately in the time
window $(0,\pi/3A)$}
using HF, NEGF and tDMRG.  The results for the TD occupation of the 
LUMO  are shown in 
Fig.~\ref{fig:density_tdmrg} (bottom panel)
and we see that the NEGF dynamics agrees very well with the tDMRG
one. The
HF results on the other hand are qualitatively different,
and show almost no dynamics for the LUMO density.

%
%
%

\section{Donor-C$_{60}$ real-time dynamics}\label{sec:nonequilibrium}
In this Section we discuss real-time simulations of the electron 
dynamics of the full donor-C$_{60}$ system driven out of equilibrium 
by the external field in Eq.~(\ref{extham}) switched on between $t=0$ 
and {\color{black} approximately $t=\pi/(3A)$}. This corresponds to a
duration of about 3~fs {\color{black} and a pumping of about one electron from HOMO 
to LUMO}. 
The NEGF results are shown in Fig.~\ref{fig:densities_c60_2b} and 
Fig.~\ref{fig:densities_c60_hf}, where HF calculations are also 
reported for comparison.        
In Fig.~\ref{fig:densities_c60_2b} we display three snapshots of the 
excess electron density \begin{align}\label{excessn}   
\Delta n({\bf r},t)= 
\sum_{ij}[\rho_{ij}(t)-\rho_{ij}(0)]\varphi_{i}^{\ast}({\bf
r})\varphi_{j}({\bf r})
\end{align}
where $\varphi_{i}$ is the $p_{z}$-orbital of carbon $i$. Compared to
the ground
(initial) state density at $t=0$, large orange (blue) regions
indicate regions of higher (lower) electron density at time $t$, i.e.
$\Delta n({\bf r},t) > 0$ ($\Delta n({\bf r},t) < 0$).

In Figs.~\ref{fig:densities_c60_hf}c and ~\ref{fig:densities_c60_hf}d
we show the excess density {\em per shell}, where the shells are
defined according to their proximity to the LUMO (hence the first
shell consists only of atom 1). For better visualization each shell
is given a different color. The shell geometric position is indicated
in the small C$_{60}$ model in the Figure.

We now go on to describe the excitation process. The action of the 
external field (photoexcitation) fills the LUMO level of the donor 
during the first 3 fs (as seen in 
Fig.~\ref{fig:densities_c60_hf}{\color{black}{a--b}}). 
This coincides with a depletion of electrons in the shells of the 
C$_{60}$ closest to the donor atom as seen in 
Fig.~\ref{fig:densities_c60_2b}. This is an image charge effect 
reflecting the increased interaction felt by the electrons on the 
C$_{60}$ due to the additional LUMO charge. Once the pumping is over 
the charge on the LUMO flows into the C$_{60}$, thus forming a CT 
exciton (excess of charge at the interface shells). This is indicated 
by the minimum in the LUMO occupation at 9.8 fs for NEGF and 6.5 fs 
for HF calculations (see again 
Fig.~\ref{fig:densities_c60_hf}{\color{black}{a--b}}) and the 
corresponding charge distribution change in the C$_{60}$ is shown in 
the middle images of Fig. 3.  
During this action of the external field and shortly after the end of
it, i.e., up until the formation of the CT exciton, the charge
dynamics in TD-HF has many features in common with NEGF,  see
Fig.~\ref{fig:densities_c60_hf}. However, at this point the two
models strongly diverge.

In the case of NEGF, the charge spreads over the entire molecule and
stabilizes on the central shells, as it can be seen in
Fig.~\ref{fig:densities_c60_hf}c. Even though there are still
oscillations in the charge density at the end of the simulation, the
average charge on the LUMO is between 0.5-0.3 -- much less than unity
(see Fig.~\ref{fig:densities_c60_hf}a). This indicates that the CT
exciton has partially decayed into a CS state.
Interestingly, we note that the majority of charge is transferred in
less than 10 fs and for all model parameters the density has
stabilised within $50$ fs. 

In contrast, in TD-HF the system does not evolve toward a CS state: 
the LUMO occupation periodically returns to the value reached just 
after the photoexcitation, as seen in 
Fig.~\ref{fig:densities_c60_hf}{\color{black}{b}}. From 
Fig.~\ref{fig:densities_c60_hf}d we see that the charge oscillates 
back and forth through the molecule with a period slightly longer 
than $\simeq 10$ fs. The difference can also quite strikingly be seen 
in Fig.~\ref{fig:densities_c60_2b} where,           
after 52 fs, the TD-HF distribution has returned to a situation very
closely resembling that after 3 fs and the NEGF distribution has
settled in a new state.

Similar scenarios are observed for a range of parameters around the 
values introduced in Section~\ref{sec:ground_state}, see 
Fig.~\ref{fig:densities_c60_hf}(a-b). Although the amount of charge
transferred from the donor to the C$_{60}$ may vary, 
the system always evolves
toward a CS state in NEGF whereas it bounces back to the initial 
state in TD-HF. 

The results presented in this Section point to the possibility of 
obtaining a CS state driven by correlation-induced decoherence. This 
process would occurr in a few tens of fs.
Experimental indications of such ultrafast processes have been 
observed~\cite{Marsh10,Dowgiallo14}, although the explanations were
somewhat different.
In the NEGF language, capturing a mechanism based on
correlation-induced decoherence 
requires a space-time nonlocal self-energy. It is therefore out of
reach of TD-HF or any other time-local theory like e.g. TD-DFT with
adiabatic functionals.

\section{Electron-phonon dynamics}\label{sec:nuclear}
We here address the effects of a single nuclear vibration 
of the Holstein~\cite{Holstein59} (local) type coupled
to the LUMO density as in Eq.~(\ref{eq:ham_eph}). 
The photo-induced oscillations of the coordinate $x$ 
renormalizes the energy levels, in agreement with the physical
picture  
obtained using TD--DFT.\cite{Falke14,Rozzi13}  

\begin{figure}[tbp]
\includegraphics[width=\columnwidth]{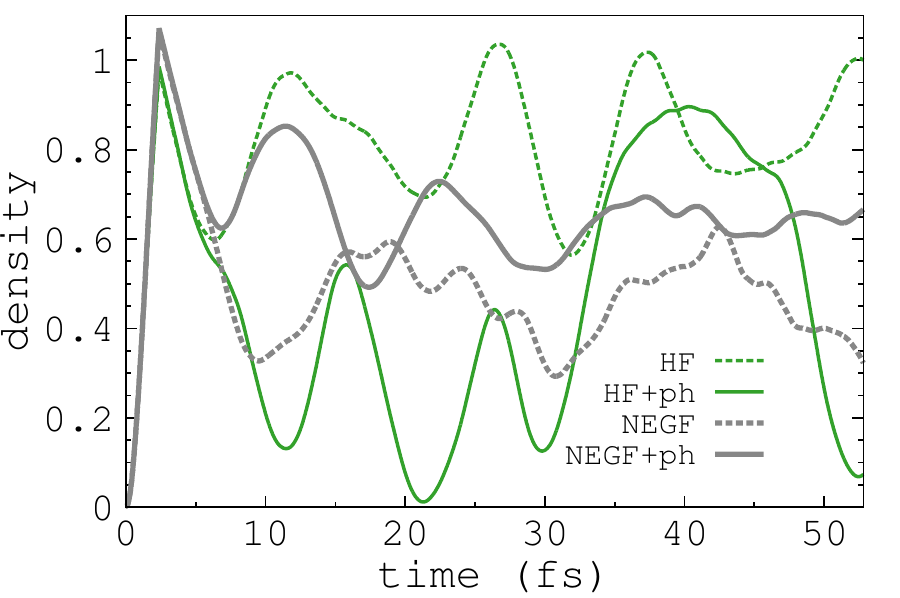}
\includegraphics[width=\columnwidth]{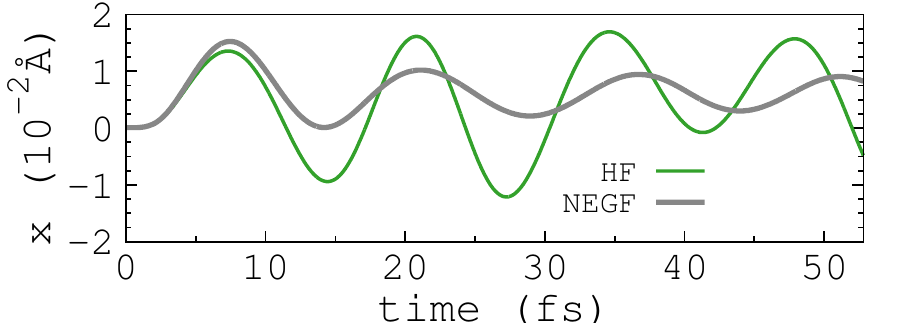}
\caption{\small LUMO occupation (top) and nuclear coordinate (bottom) 
for a donor-C$_{60}$ system with electron-phonon coupling.
{\color{black} In the initial state the LUMO and C$_{60}$ levels are
partially aligned.} Results 
are shown for HF (green-solid) and NEGF (grey-solid). For better
comparison we 
also show results without electron-phonon coupling (dashed).}
 \label{fig:nuclear_f1}
\end{figure}

We show in Fig.~\ref{fig:nuclear_f1} results for the same parameters
as in Figs.~\ref{fig:densities_c60_2b} and \ref{fig:densities_c60_hf}
with $\omega_{ph} = 0.548$~eV, $g=20$~eV/\AA~and $M$ about forty 
times the proton mass. The NEGF dynamics
is largely unaffected
by the nuclear vibration, the only change being a slight decrease in
the charge transferred. The Holstein mode changes quantitatively the 
TD-HF behavior but there is still no clear evidence of CS. 

\begin{figure}[tpb]
 \includegraphics[width=\columnwidth]{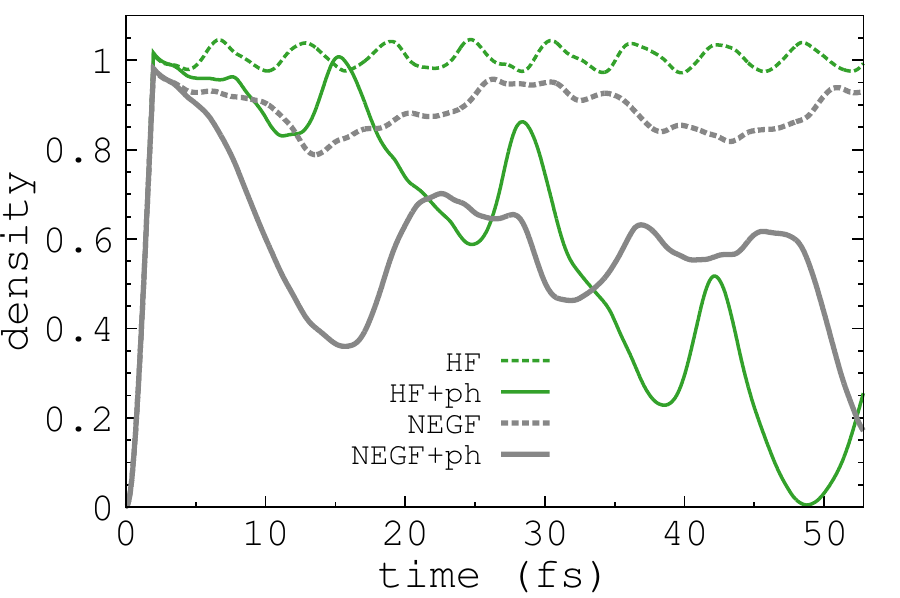}
 \includegraphics[width=\columnwidth]{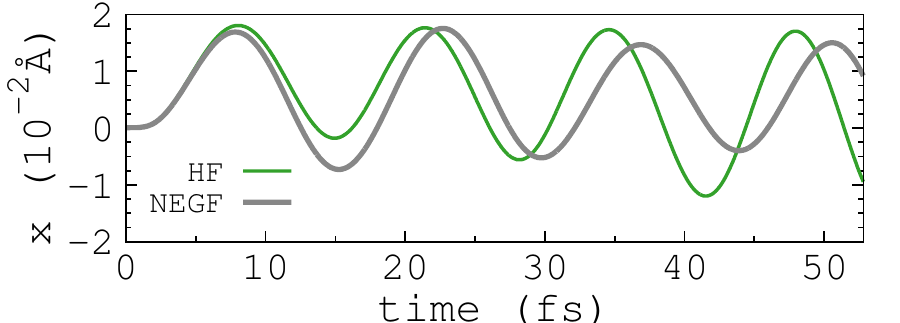}
 \caption{\small LUMO occupation (top) and nuclear coordinate
(bottom) 
for a donor-C$_{60}$ system with electron-phonon coupling and 
modified donor parameters, see main text. {\color{black} In the initial
state the LUMO and C$_{60}$ levels are misaligned.} Results 
are shown for HF (green-solid) and NEGF (grey-solid). For better
comparison we 
also show results without electron-phonon coupling (dashed).}
 \label{fig:nuclear_f2}
\end{figure}

It is interesting to explore a different regime where, at clamped 
nuclei, due to level misalignment there is no CT 
either in NEGF nor in TD-HF. For this purpose we  
change the donor parameters and consider $\epsilon_H=-1.4$~eV, 
$\epsilon_L=1.6$~eV, $U_{HL}=0.6$~eV and $U_{La}=2U_{Ha}=1.2$~eV.
The results are shown by the dashed curves in
Fig.~\ref{fig:nuclear_f2}. When the coupling to the nuclear vibration
is turned on, with $\omega_{ph}$, $g$ and $M$ as before, 
a dramatic  change in the CT behavior is observed. 
The TD-HF calculations go from
showing almost vanishing CT to efficient CS. A similar regime with 
analogous findings has 
been investigated by other authors using adiabatic 
TD--DFT~\cite{Falke14,Rozzi13}. We also notice that space-time 
nonlocal electronic correlations (NEGF results) do not change the
qualitative picture, 
the main effect being a less efficient CS. This can easily be
understood in terms of quasi-particles. In HF the quasi-particles
have an 
infinitely long life-time or, equivalently, a sharp energy.
Therefore, when the 
phonon-shifted HF LUMO energy aligns to an unoccupied 
acceptor level the CT is extremely efficient. 
{\color{black}{Beyond HF the quasiparticle weight spread 
over several energies and CT occurs  whenever the donor and 
acceptor spectral densities overlap. As this overlap is always 
smaller than unity only a fraction of the 
electron charge can be transferred.}}

The reported evidence is based on simulations
with Holstein-like phonons. However, we performed calculations
with Fr\"ohlich-like phonons as well (not shown), i.e., we replaced
the coupling 
$g\,\hat{n}_{L}x$ with 
$g\sum_{\sigma}(\hat{c}^{\dag}_{L\sigma}\hat{c}_{1\sigma}+h.c.)x$,
arriving at similar conclusions.
\section{Conclusions and outlook}\label{Conclusions}
We performed real-time NEGF simulations of the charge dynamics in a 
donor-C$_{60}$ dyad photoexcited by external laser pulses.
The accuracy of the NEGF scheme was preliminary assessed through 
comparison with tDMRG data  in D--A one-dimensional 
model systems, finding excellent agreement.

We reported results for two different physical situations, namely
levels 
alignment allowing and forbidding the formation of a CT exciton. In 
the former case we find that the CT exciton dissociates and the 
system evolves toward a CS state even with clamped nuclei. We 
highlighted the role of space-time nonlocal correlations (or, 
equivalently, correlation-induced decoherence) by comparing NEGF and
TD-HF. In 
TD-HF the CT exciton recombines and no evidence of CS is observed. 
We also verified that the inclusion of nuclear motion does not change
the picture.
However, in the second physical situation, i.e., level alignment 
forbidding CT, we find that the impact of 
nuclear motion is substantial. In fact, both TD-HF and NEGF predict 
CS due to coupling to nuclear vibrations, 
albeit in NEGF the yield of exciton dissociation is smaller. The 
TD-HF 
results agree with similar findings in organic D--A complexes 
obtained using TD--DFT.~\cite{Rozzi13,Falke14}

Although we focused on a C$_{60}$ acceptor, the NEGF method can also
be used to deal with donors and adsorbates on surfaces, as it has
been already demonstrated in the context of molecular
electronics.~\cite{SRHT.2011,JSMST.2013} In these instances, the only
approximation is that one should cut off the Coulomb interaction in
the self-energy diagrams to a finite region in the neighbourhood of
the donor or the adsorbate species.
In a broader perspective, our results put forward NEGF as a well
suited tool for an accurate description of general time-dependent
electronic phenomena, in particular to characterise how nonlocal
correlations influence realistic nanostructured systems and surfaces, 
also when nuclear degrees of freedom are involved. This might be
particularly
 important for studies of electron dynamics in the first
hundred femtoseconds after excitation -- an area which has become
accessible to experimental studies in recent years, and which has a
strong influence on the dynamical properties of for example nanoscale
solar harvesting systems, optoelectronic components, surface-induced
molecular dissociation/synthesis and desorption
phenomena.~\cite{BMV.2016} 

\section{Acknowledgements}
A. M. was supported by the Swedish Research Council (VR). 
E.P. acknowledges funding from the European Union project MaX 
Materials design at the eXascale H2020-EINFRA-2015-1, Grant Agreement 
No. 676598 and Nanoscience Foundries and Fine Analysis-Europe 
H2020-INFRAIA-2014-2015, Grant Agreement No. 654360.   
G.S. acknowledges funding by MIUR FIRB Grant No. RBFR12SW0J and EC
funding through the
RISE Co-ExAN (GA644076).

\bibliography{references}{}

\end{document}